\documentstyle[12pt]{article}
\def\eq{\begin{equation}} 
\def\en{\end{equation}}
\newcommand \be  {\begin{equation}}
\newcommand \bea {\begin{eqnarray} \nonumber }
\newcommand \ee  {\end{equation}}
\newcommand \eea {\end{eqnarray}}

\def \bi{\bibitem}

\def\d{{\rm d}}
 \def\(({\left(}
 \def\)){\right)}
\def\bi{\bibitem}

\def \ov{\over}

\def \d{{\rm d}}

\def \beqna{\begin{eqnarray}}
\def \eeqna{\end{eqnarray}}
\def \beq{\begin{equation}}
\def \eeq{\end{equation}}
\def \be{\begin{equation}}
\def \ee{\end{equation}}

\def \ov{\over}

\def \la{\langle}
\def \ra{\rangle}

\def \ab2{\alpha\beta^2}

\def \la{\langle}
\def \ra{\rangle}

\begin{document}   

\baselineskip25pt

\begin{center}

{\LARGE\bf Fluctuation Dissipation Ratio in\\
       Three-Dimensional Spin Glasses}
\vskip1cm

\end{center}

\baselineskip20pt

\begin{center}

{\Large Silvio Franz (1) and Heiko Rieger (2)}
\vskip1cm

{\large
(1)  NORDITA,\\
     Blegdamsvej 17,\\
     DK-2100 Copenhagen \O\\
     Denmark\\ }

\vskip0.5cm
{\large
(2)  Institut f\"ur Theoretische Physik,\\
     Universit\"at zu K\"oln,\\
     50937 K\"oln,\\
     Germany\\ }
\vskip0.5cm

{\large August 1994}
\vskip1cm

\end{center}

\baselineskip16pt

\vskip1cm

{\bf Abstract:} We present an analysis of the data on aging in the
three-dimensional Edwards Anderson spin glass model with nearest
neighbor interactions, which is well suited for the comparison with a
recently developed dynamical mean field theory. We measure the
parameter $x(q)$ describing the violation of the relation among
correlation and response function implied by the fluctuation
dissipation theorem.

\vfill
\eject

On experimental time scales spin glasses are out of equilibrium.
Experiments have pointed out that `aging effects', i.e. the dependence of
some measurable quantities on the time spent in the low temperature
phase after a quench, persist at least for times of the order of years
\cite{agerev}. The same kind of phemomena have been recorded in numerical 
simulations of various spin glass models (see \cite{numrev} for a review).

A lot of activity has been devoted to understand the origin of these
phenomena with phenomenological  approaches\cite{kophil,fishus,sibani89,bouch}
and from the analysis of mean field models \cite{cuku1,frme,cuku2}.
This last approach is rapidly evolving, and major progress has been
made towards a mean field theory of the off-equilibrium dynamics of
spin glasses. In this note we want to compare some features
of the mean field theory of spin glasses with the more realistic
three-dimensional Edward Anderson model. In mean
field theory (MFT) aging is associated with a phase transition, there is a high
temperature phase in which the systems equilibrate, at low temperature
phase where the systems settle in an asymptotic off-equilibrium state.

On the time scales we can reach we can not certainly claim that the
system has  reached an asymptotic behaviour, neither we can exclude a
crossover from aging to equilibrium dynamics for very long times. The
question whether aging in a 3D system is really asymptotic or gradually
disappears, although fundamental in principle,  may not be the most
relevant one from an experimental point of view.

We compare the behaviour of finite
time in 3D systems with the predictions of the mean field theory.
Among these, we will focus on a systematic analysis of the violations
of the fluctuation dissipation theorem (FDT). The FDT cannot hold in
a non-equilibrium situation, where the probability distribution for
the spin configurations is  time-dependent (see the discussion in
\cite{sibani}). The fundamentally new idea
developped in \cite{cuku1,frme,cuku2} is that a quantitative 
analysis of this violation could reveal a deeper insight into
long time off-equilibrium properties of spin glasses.

In spin glass dynamics  crucial quantities of interest are
the spin autocorrelation function $C(t,s)$ and its associated 
response function $G(t,s)$
\be
\begin{array}{ccc}
C(t,s) & = & [\la S_i(t) S_i(s)\ra]_{\rm av}\;,\\
 & & \\
r(t,s) & = & \partial \la S_i(t)\ra/\partial h_i(s) \quad (t>s)\;,\\
\end{array}
\ee
where $\la\cdots\ra$ means an average over the stochastic process
describing the dynamical evolution of the system at a temperature
$T=\beta^{-1}$ (starting with a random initial configuration) and
$[\cdots]_{\rm av}$ means an average over the quenched disorder. At
thermal equilibrium these function are homogeneous, and related by the
fluctuation dissipation theorem relation $r_{eq}(t-s)=\beta\partial
C_{\rm eq}(t-s)/\partial s$. In general, to characterize off-equilibrium
situations it is possible to introduce the `fluctuation dissipation
ratio', as the function
\be 
x(t,s)= {r(t,s)\ov\beta\partial C(t,s)/\partial s}.
\ee
It is convenient for the following analysis to change a bit the definition of 
this function.
First we  define a
function $\tilde{s}(q,t)$ as the time $s$ such that $C(t,s)=q$, which
is unique due to the monotonicity of $C(t,s)$ with $s$. Then
we consider the fluctuation-dissipation-ratio at this time
\be
\overline{x}(t,q)=x(t,\tilde{s}(q,t))\;,
\label{trafo}
\ee

The above mentioned MFT makes a particular set of predictions for 
$x(t,s)$ (and {\it a fortiori} for $\overline{x}(t,q)$)
in the limit $t,s\to \infty$. There are different ways in which
one can take this limit, depending on the relation among $t$ and $s$.
In ordinary equilibrating systems, the relevant procedure is
to fix the difference $t-s=\tau$ to a finite value.
This yields to limiting functions $C_{as}(\tau)$, $r_{as}(\tau)$.
The correlation function
$C_{as}(\tau)$ decreases monotonically from the value 1 at $\tau=0$ to 
a value that we call 
$q_{EA}$ for $\tau\to\infty$, and the FD relation is respected ($x=1$). 
In any different 
limiting procedure, that would imply $t-s\to\infty$, one would find 
that the correlation function tends to $q_{EA}$. In other words, for 
$t,s\to\infty$ all the observable dynamical effect are concentrated 
in the finite $\tau$ region. 
In aging systems this does not happen: dynamical effects
persist in regions of the plane $(t,s)$ where the limit is taken differently.
This is appearent in experiments \cite{agerev}
 where important dynamical effecs 
are observed on time scales $\tau$ of the order of the `waiting time'
($s$). 

In mean field spin glasses, 
dynamics take place both in a region of time homogeinity where the FDT relation
is respected,  
and in an aging region.
This was first theorized in \cite{cuku1},
and then verified by a numerical solution of mean field off-equilibrium
dynamical equations for a particular model in \cite{frme}. An ansatz
which allows for a precise definition of the infinite time limit 
in the homogeneous and  aging regimes,  has been put forward
in \cite{cuku1,frme,cuku2}. Without entering in the details of this limiting
procedure we just reassume some consequences of the analysis. 
The time homogeneous regime is qualitatively
similar to an equilibrium regime where 
$\lim_{\tau\to\infty}C_{as}(\tau)=q_{EA}$, 
$x(q)=1$ for $q_{EA}< q\leq 1$.

In the aging regime the function $C(t,s)$ decreases, for decreasing $s$,
from $q_{EA}$ to a value $q_{min}$ ($q_{min}=0$ for spin glasses in
absence of a magnetic field). 
The function $\overline{x}(t,q)$ tends, 
for any $q$ in the interval $[q_{min},q_{EA}]$, to a well defined limit
$x(q)$.

It turns out that the function $x(q)$ is formally related to the
inverse of the {\it static} Parisi function $q_{\rm stat}(x)$, $x_{stat}(q)$.  A
non trivial $x(q)$ is found in these models which statically exhibit
replica symmetry breaking \cite{mpv}.
In all cases in which the replica symmetry
breaking is associated with a continuos $q_{\rm stat}(x)$, then
$x_{\rm stat}(q)=x(q)$. If the static $q_{\rm stat}(x)$ is discontiuous 
 $x(q)$ turns
out to be different from its static couterpart. This is the case e.g.
of the p-spin spherical model \cite{cuku1}, where $x(q)$ is a step function.
However in both cases, $\d x(q)\ov \d q$ has all the properties
defining a probability distribution, as it happens for 
${\d x_{\rm stat}\ov \d q}$. At present there is not a
complete physical comprehension of the relation among the static
definition of $x(q)$ and the dynamic one, and of the fact that the
latter is associated to a probability distribution.

In this paper we want to try to extract the above defined function
$x(q)$ from numerical data obtained for the 3D
Edwards-Anderson model via Monte-Carlo simulations performed by one of
us recently \cite{easim}. The advantage of numerical simulations
compared to experiments is that while experimentally it is very
difficult to get direct information on the correlation function, 
 in numerical simulations one
can easily have access both to the correlation and the response
functions. 

The correlation function is measured directly in the course of
simulations starting from a random initial condition, which
corresponds to a rapid temperature quench from the paramagnetic phase.
The response
function is measured in `TRM (thermoremanent magnetization) 
experiments': the system is let to age
for a time $t_w$ in presence of a small magnetic field $h$, then the
magnetic field is cut off and the magnetization is recorded as a
function of the time $\tau$ measured starting from $t_w$. We assume
linear response conditions where the magnetization
$M(\tau+t_w,t_w)$\footnote{
Note that we use a notaton in which both the time arguments of $M$ are measured starting from the quenching time $t=0$. The standard notation would be 
$M(\tau,t_w)$.}
 is given by:
\be 
M(\tau+t_w,t_w)={h} \int_0^{t_w} r(\tau+t_w, s)\;\d s.
\label{mtrm}
\ee 
Using (\ref{trafo}) one can write
\be
M(\tau+t_w,t_w)={h\ov T}\int_{C(\tau+t_w,0)}^{C(\tau+t_w,t_w)}
\overline{x}(\tau+t_w, q)\;\d q
\ee
and exploiting the monotonicity of $C$ this time with respect to
$\tau$, we choose $\tau$ such that $C(\tau+t_w,t_w)=q$ and
write with obvious meaning of the symbols
\be 
M(q,t_w)={h\ov T}\int_{C(\tau+t_w,0)}^q 
\overline{x}(\tilde{\tau}(q,t_w)+t_w, q')\;\d q'
\ee
For infinite $t_w$, assuming loss of memory of the initial condition,
$\lim_{t_w\to\infty}C(\tau+t_w,0)\to 0$,  
one would have $M(q)={h\ov T}\int_{0}^q \d q' x(q')$.
In the following we will present simulation data for the function
\be
\chi(q, t_w)=\frac{T}{h} M(q,t_w)\stackrel{t_w\to\infty}{\longrightarrow}
\chi(q)
\ee
in the 3D Edward Anderson model.
Simulation data for the corresponding function in the Sherrington and
Kirkpatrick model have been given in \cite{cuku2}. In order to
understand our findings let us discuss some simple scenario for the
function $\chi (q)$. 

\noindent
{\bf 1)} \underline{Ergodic behaviour in the whole phase space:}

\noindent
In this case $q_{EA}=0$ and 
$x(q)$ is equal to one in the whole interval $0\leq q\leq 1$ and 
one finds the classical FDT results 
\be
\chi(q)=q
\ee
typical e.g. of the paramagnetic systems.

\noindent
{\bf 2)} \underline{Ergodic behaviour in a confined component:}

\noindent 
Here the systems relaxes to a non zero $q_{EA}$ and the dynamics remains 
confined to a single valley, then 
$x(q)=\Theta(q-q_{EA})$ and 
\be
\chi=\Theta(q-q_{EA}) (q-q_{EA})
\label{onevalley}
\ee
Such a behaviour is found e.g. in ferromagnets in the low temperature phase, 
where $q_{EA}$ is equal to the square of the magnetization and it is would
also be the prediction for a spin glass scenario like that proposed by
Fisher and Huse \cite{fishus}.

\noindent
{\bf 3)} \underline{Mean field aging behaviour:}

\noindent
In this case two scenarios have been found in the literature.  In
models with one step of replica symmetry breaking $x(q)=x
\Theta(q_{EA}-q) +\Theta(q-q_{EA})$ and $\chi(q)= x \Theta(q_{EA}-q) q
+\Theta(q-q_{EA})\{q-(1-x)q_{EA}\}$ while in models with continuous
replica symmetry breaking $x$ is an increasing function from zero at
$q=1$ to one for $q=q_{EA}$ and stays equal to that value for
$q>q_{EA}$.  Correspondingly
\be
\chi(q)=\Theta(q_{EA}-q)\int_0^q \d q' \tilde{x}(q')
+ \Theta(q-q_{EA})\left\{q-q_{EA} +\int_0^{q_{EA}} \d q' x(q')\right\}\;. 
\ee
In  the SK model near $T_c$ 
it is found \cite{cuku2} the linear shape  $x(q)=2 a q$ with $a=1/2$ 
and  one obtains
$
\chi(q)=\Theta(q_{EA}-q)a q^2+ \Theta(q-q_{EA})(q-q_{EA} +a q_{EA}^2)\;.
$

Let us turn now to the presentation of the simulation data. We stress
that at finite times $C(t,s)$ and $r(t,s)$ are regular functions, and
the possible singularities in $x(q)$ and $\chi$ should be
smoothed in some crossover region. We use the data one of us obtained
in \cite{easim} and did some additional runs where necessary. For
completeness let us recall the definition of the model that we
investigate: it is the three-dimensional Edwards-Anderson model
defined by the Hamiltonian
\be
H = -\sum_{\langle ij\rangle} J_{ij}\sigma_i\sigma_j - h \sum_i S_i\;,
\label{hamilt}
\ee
where $\langle ij\rangle$ are nearest neighbor pairs on a simple cubic
lattice, $S_i=\pm1$ are Ising spins, $J_{ij}$ are quenched random
variables taking on the values $+1$ and $-1$ with equal probability
and $h$ is an external magnetic field.
In this model a phase transition has been observed
at $T_c\approx1.2$ \cite{tcsim} (see however \cite{Marinari} for a different
point of view). We use single-spin-flip heat-bath dynamics with
parallel sublattice update and calculate the spin-autocorrelation
function $C(\tau,t_w)$ (in zero field) and the thermoremanent
magnetization $M(\tau+t_w,t_w)$ as defined in (\ref{mtrm}). The field $h$
applied for a time $t_w$ before starting the measurement is small
($h=0.1$), and we checked by looking at $h=0.05$ and $h=0.2$, too, that
we are in the lineare response regime. Thus 
$\chi(\tau+t_w,t_w)=T/h\,M(\tau+t_w,t_w)$ is the magnetic relaxation function
occuring in linear response theory. The lattice size used is
$N=32^3$ and we made sure that finite size effects were not significant.
All data are averaged over at least 128 samples (i.e.\ different
realizations of the disorder).

In figure 1 we show a picture, analogous to that presented in \cite{joa},
that clearly show the violation of the FDT relation among magnetization and 
correlation function as a function of time. As long as $\tau\ll t_w$
the FDT-relation is fulfilled
\be
\chi_{\rm dc}-\chi(\tau+t_w,t_w)=\beta\{1-C(\tau+t_w,t_w)\}\;,
\ee
where $\chi_{\rm dc}$ is the equilibrium dc-susceptibility (see e.g.\
\cite{kophil}). For $\tau\gg t_w$ this relation is obviously violated.

In figure 2 we present the function $\chi(q,t_w)$ for different
waiting times and temperatures $T=0.8,1,1.5,2$.  It is clearly seen in
the $T=2.0$ plot, that after a short transient $\chi$ tends to the
paramagnetic function $\chi(q)=q$.  In the plot of the $T=1.5$ data we
can see that the system has not equilibrated even after the largest
waiting time $t_w=10^5$.  At low temperature we clearly recognize a
$t_w$-dependent linear part in $\chi$ at large $q$. The slope of the
linear part is indeed 1 as it is shown in figure 3 where we display
$\chi(q,t_w)-q$ for T=0.8.  From figure 2 and 3 one can extract an
effective time dependent EA parameter $q_{EA}(t_w)$ as the value of
$q$ at which $\chi(q)$ starts to depart from linearity. In this way we
estimate at $T=0.8$ for $t_w=10^3$, $t_w=10^4$, $t_w=10^5$ the values
$q_{EA}=0.78,0.75,0.72$, respectively. It is clear that these
data do not allow for any extrapolation.

The small $q$ part of the curves can be reasonably fitted with an arc
of parabola $\chi(q)=a q^2$, for $t_w=10^3,10^4,10^5$ the value of $a$
at $T=0.8$ is roughly constant and equal to $a=0.2$.  A linear fit of
the kind $\chi(q)= x q$ gives much poorer results.  This seems to
indicate a scenario more similar to that of SK-like continuous replica
symmetry breaking than that of a one-step replica symmetry breaking.

In figure 4 finally we present $\chi(q)$ as a function of $q$ for
$t_w=10^5$ and different temperatures.  As expected the apparent
$q_{EA}$ parameter grows for decreasing temperatures. On one side one
definitely still observes a slight dependence of $\chi(q,t_w)$ on the
waiting time $t_w$, which means that rigorous statements on the
limiting shape of $\chi(q)$ and hence of $x(q)$ hardly can be
made. On the other side we do not observe any tendency of the curves
$\chi(q,t_w)$ to approach a form like (\ref{onevalley}) that is
characteristic for a system with only two pure states (note that this
would imply that the whole small-$q$-part, i.e.\ $q<q_{EA}$, of
$\chi(q,t_w)$ has to come down to zero). 

We leave to the reader to judge if our data 
can be interpreted as an
indication for a nontrivial $x(q)$ 
in three dimensions. However, the 3D
EA-model is known to be only marginally critical, therefore it would
be highly desirable to perform the same kind of investigation 
in four dimensions, where a nontrivial {\it static} $P(q)$ has already been 
reported from a finite-size scaling analysis \cite{reger}.

Concluding we have analized in this paper the data for the correlation
and the response functions in the light of a recent mean field theory
of aging phenomena.  We have shown that at least as the quantity
$\chi(q)$ is concerned, the behaviour of the 3D EA model at the time
scale we investigate agrees qualitatively with a mean field like
behaviour. One clearly sees a separation of the dynamics in a
quasi-equilibrium part, analogous to an equilibrium dynamics where the
FD relation is respected, and an aging part where the FD ratio take
values different from zero and one. Rough estimates indicate that
$x(q)$ grows linearly with $q$ for small $q$, a behaviour reminiscent
of the SK model.  The time scales to which we have access prevent us
to probe the asymptotic behaviour of the system, and even to prove
that aging phenomena do not gradually disappear for increasing waiting
times.  This question is related the one longly debated of the
existence of a sharp phase transition in the model, and more general
in 3D short range spin glasses. Although of fundamental theoretical
importance, due to the slowness of the relaxation process, it is
certainly not the most interesting one from an experimental point of view.
It could well be the case that even if the transition is absent and
the aging is interrupted after some very long time, the mechanisms
responsible for aging in mean field could be relevant for the 3D
physics on experimental times.

HR's work was performed with the Sonderforschungsbereich 341 
K\"oln--Aachen--J\"ulich supported by the DFG. 
He also thanks the NORDITA institute for its kind hospitality.

\vfill
\eject

\vfill
\eject

{\large\bf Figures}
\vskip1cm
\begin{itemize}
\item[Fig.\ 1]
The quantities $\chi_{dc}-\chi(\tau+t_w,t_w)$ $(\circ)$ 
and $\beta[1-C(\tau+t_w,t_w)]$ $(\bullet)$ versus
time $t$ for different waiting times. $\chi_{dc}$, the dc-susceptibility 
is a single fit paramater for all waiting times. The temperature is $T=0.7$.
Note that for the FDT to hold both curves have to be identical.

\item[Fig.\ 2]
The function $\chi(q,t_w)$ versus $q$ for various temperatures.
The waiting times are $t_w=10^2$ ($\circ$), $t_w=10^3$ ($\triangle$), 
$t_w=10^4$ ($\Box$) and $t_w=10^5$ ($\bullet$).

\item[Fig.\ 3]
The function $q-\chi(q,t_w)$ versus q for $T=0.8$ and different waiting times.
$q-\chi(q,t_w)$ should be constant as long as the FDT is fulfilled.
The full line is only a guide for the eye.

\item[Fig.\ 4]
$\chi(q,t_w)$ versus $q$ for different temperatures at $t_w=10^5$.

\end{itemize}

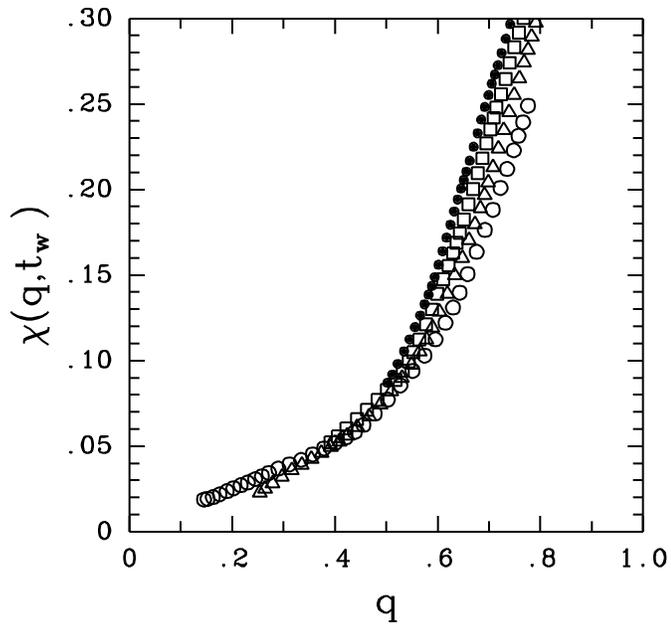
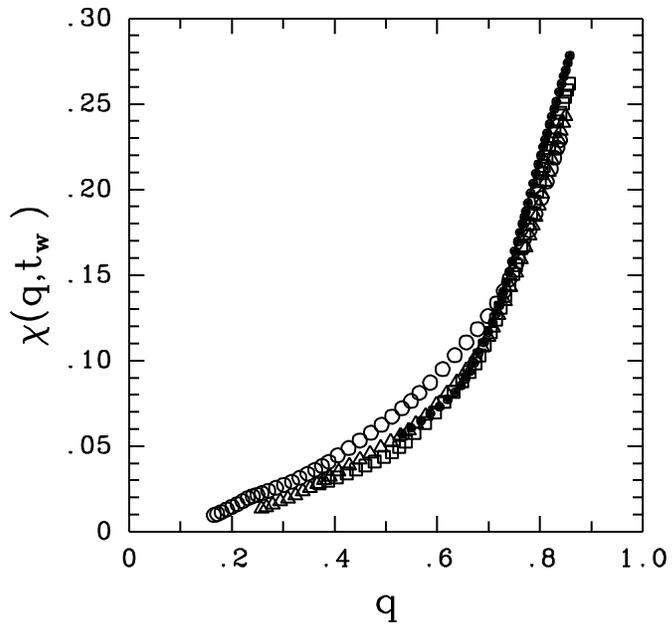
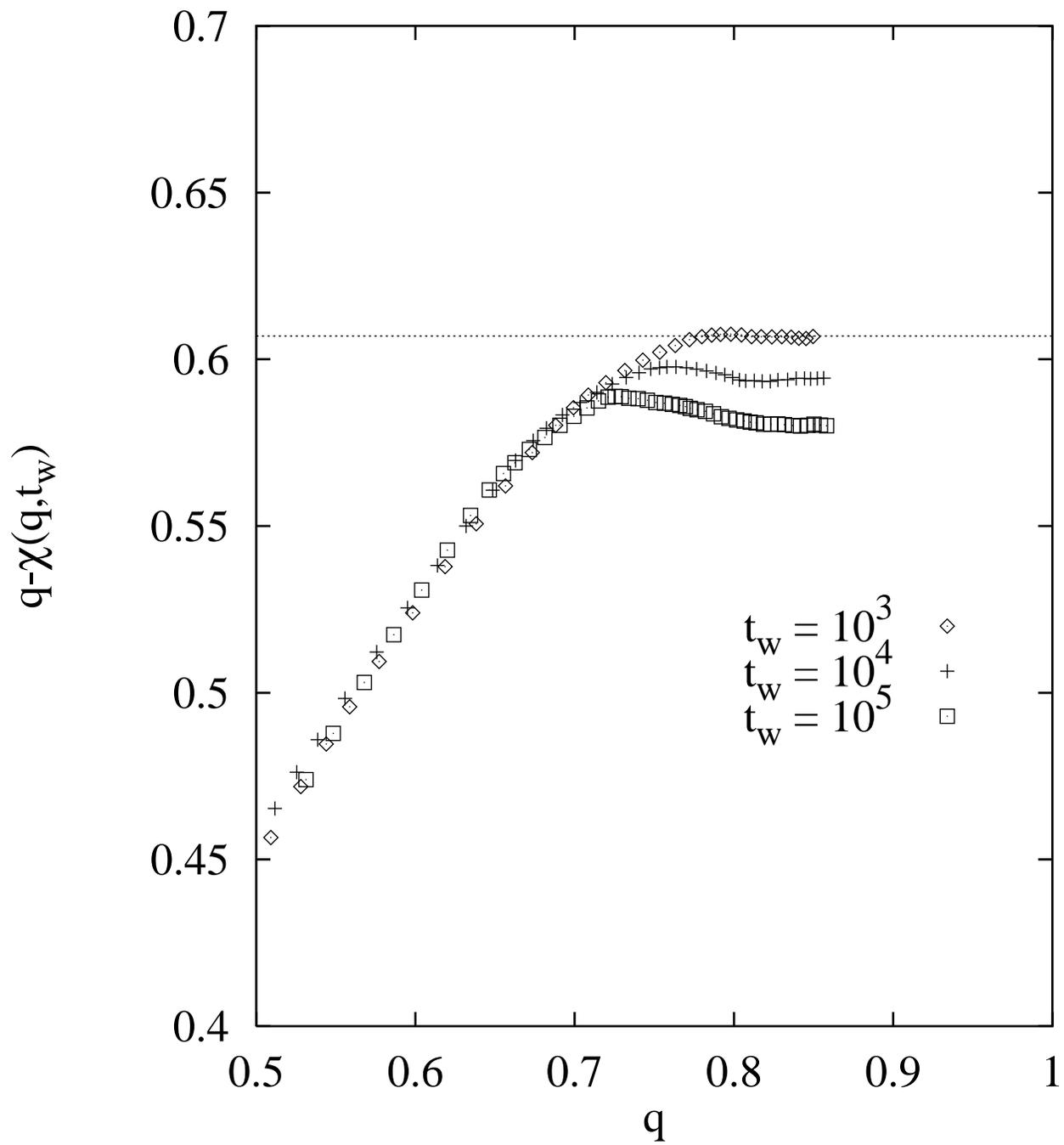
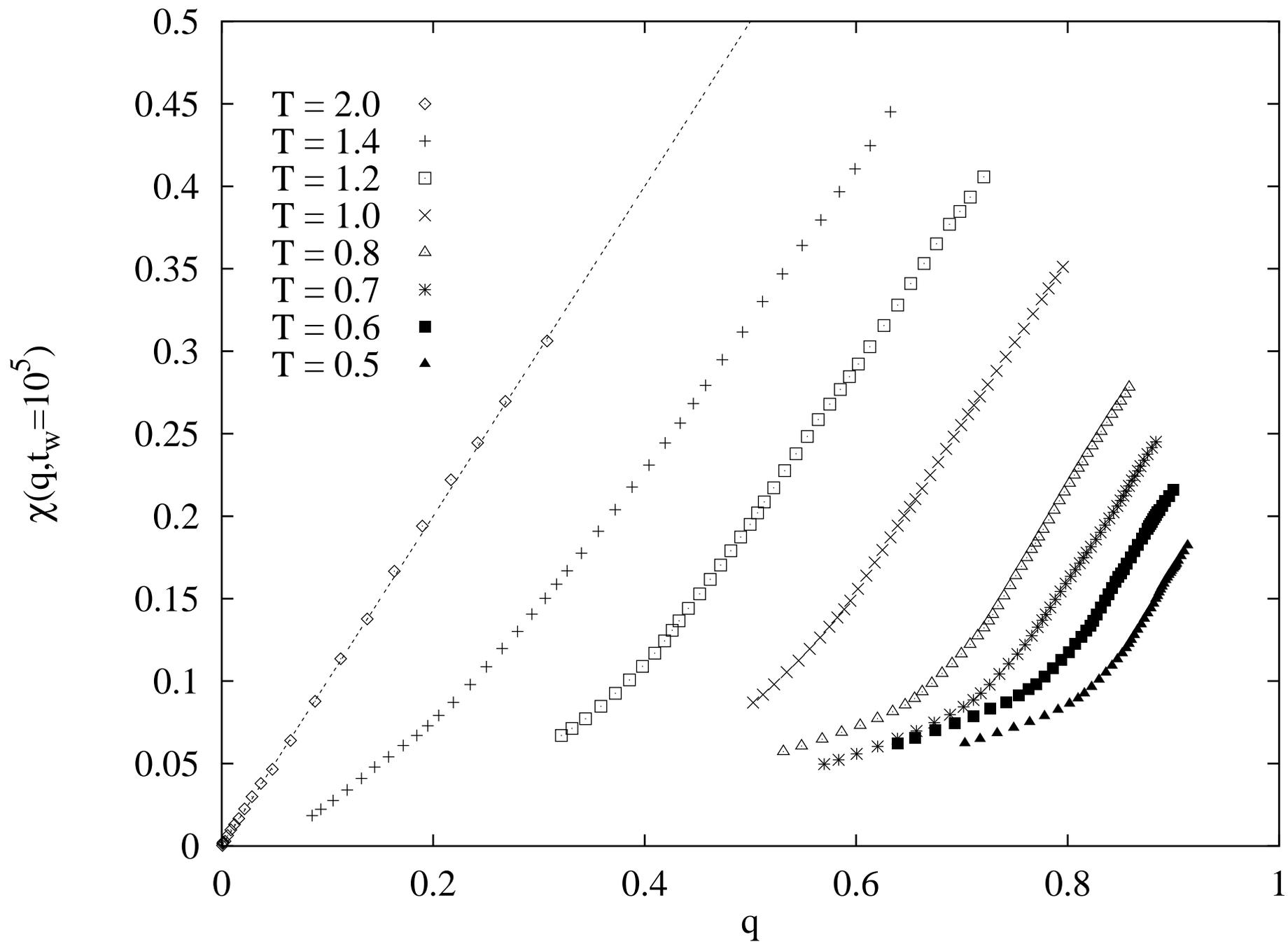
\end{document}